\documentclass[11pt]{article}
\usepackage[margin=1in]{geometry}
\usepackage{amsmath,amssymb,amsfonts,bm}
\usepackage{authblk}
\usepackage{hyperref}
\usepackage{physics}
\usepackage{graphicx}
\usepackage{minted}
\usepackage{xcolor}
\usepackage{indentfirst}
\usepackage[numbers,sort&compress]{natbib}

\title{Learning the Inverse Ryu--Takayanagi Formula with Transformers}

\author[]{Sejin Kim\thanks{ 
\href{mailto:sejin@kias.re.kr}{sejin@kias.re.kr}}}

\affil[]{Center for Artificial Intelligence and Natural Sciences, Korea Institute for Advanced Study, Seoul 02455, South Korea}
\date{}

\begin{document}
\maketitle

\begin{abstract}
We study the inverse problem of holographic entanglement entropy in AdS$_3$ using a data-driven generative model. Training data consist of randomly generated geometries and their holographic entanglement entropies using the Ryu--Takayanagi formula. After training, the Transformer reconstructs the blackening function within our metric ansatz from previously unseen inputs. The Transformer achieves accurate reconstructions on smooth black hole geometries and extrapolates to horizonless backgrounds. We describe the architecture and data generation process, and we quantify accuracy on both $f(z)$ and the reconstructed $S(\ell)$. Code and evaluation scripts are available at the provided repository.

\end{abstract}

\section{Introduction}

The AdS/CFT correspondence relates gravity in asymptotically AdS spaces to conformal field theories (CFTs) on their boundaries \cite{Maldacena:1997re, Witten:1998aa,Ramallo:2013aa,Ryu:2006bv}. Entanglement entropy provides a bridge between the two sides through the proposal of Ryu and Takayanagi, which expresses boundary entanglement in terms of the area of a minimal surface in the bulk \cite{Ryu:2006ef,Hubeny:2007aa,Casini:2011aa,Headrick:2007aa,Lewkowycz:2013aa,Camps:2013aa,Ryu:2006bv,Ji:2025aa}. 

We ask whether, given entanglement entropy in an AdS$_3$ background, one can reconstruct the blackening function that characterizes the metric \cite{Hamilton:2006aa,Bilson:2008aa,Bao:2019aa,Bilson:2011aa,Czech:2015ab,Dong:2016aa,Park:2023aa,Ji:2025aa,Jokela:2025aa,Ahn:2024aa,Kim:2024aa,Deb:2025aa}. Using the Hamilton--Jacobi equation, which provides a link between the boundary size $\ell$ and the turning point $z_t$ of the minimal surface, we recast the problem as learning a map from $\ell(z_t)$ to the blackening function $f(z)$. We impose asymptotic AdS boundary behavior and focus on regular black hole geometries with a single horizon. This work focuses on approximating the inverse relation defined by the Ryu--Takayanagi formula within our metric ansatz. We do not attempt to learn the full solution space of the gravitational field equations. Related reconstruction problems have also been studied from a complementary perspective in the context of pole-skipping, where the near-horizon geometry can be recovered analytically from discrete momentum-space pole-skipping points of boundary Green’s functions \cite{Lu:2025aa,Lu:2025ab}. In contrast, the present work focuses on a data-driven approximation to the inverse Ryu--Takayanagi map in AdS$_3$ using entanglement entropy as input.

In previous studies, the target function was represented by a neural network, and the loss function was constructed as a weighted sum of the equations of motion and several boundary conditions \cite{Yaraie:2021aa,Hashimoto:2018wm,Hashimoto:2018aa,Song:2020wm,Akutagawa:2020yeo,Park:2022aa,Ahn:2024aa,Ahn:2025aa,Lee:2025aa,Filev:2025aa,Deb:2025aa,Kim:2024aa,Hashimoto:2024aa,Bea:2024aa,Ahn:2024ab}. Such approaches can work well for a single instance, but the loss must be redesigned for each theory choice. In contrast, we take a data-driven approach. We generate pairs of boundary and bulk quantities from the Ryu--Takayanagi formula and we train a simple Transformer to map boundary inputs to bulk outputs. The network optimizes its own parameters rather than the physical variables so inference does not require a problem specific loss function. 

The key point of this study lies in data generation. Stochastic white noise is added to the blackening function before computing entanglement entropy. This noise sharpens the local sensitivity of $\ell(z_t)$ to $f(z)$, helping the Transformer understand the underlying integral equation.

After a single training phase, the trained Transformer reconstructs the blackening function $f(z)$ for unseen inputs that include smooth black hole geometries and, in many cases, horizonless backgrounds. We quantify accuracy both on the predicted blackening function and on the reconstructed entanglement curve, and we make code and scripts available for full reproducibility.

\section{Data-driven approach}
To date, holographic inversion has often relied on optimization-based methods \cite{Yaraie:2021aa,Hashimoto:2018wm,Hashimoto:2018aa,Song:2020wm,Akutagawa:2020yeo,Park:2022aa,Ahn:2024aa,Ahn:2025aa,Lee:2025aa,Filev:2025aa,Deb:2025aa,Kim:2024aa,Hashimoto:2024aa,Bea:2024aa,Ahn:2024ab}, including direct-search approaches \cite{Kim:2024aa} and physics-informed neural networks that represent target function as a neural network \cite{Yaraie:2021aa,Raissi:2017aa,Raissi:2017ab,Champion:2019aa}. These methods are engineered for a given instance by encoding equations of motion together with boundary and horizon conditions into the loss. They can be effective single instance solvers, yet they require problem specific design and do not transfer easily across model families.

The optimization-based loss function can be written as
\begin{equation}
  {\rm Loss}_{\mathrm{opt}}\left(\Phi, V(\Phi), \cdots ; {\rm EOM}, {\rm HC}, \cdots\right)
  = \|{\rm EOM}\|_2^{ 2} + \epsilon_1 \|{\rm HC}\|_2^{ 2} + \cdots,
\end{equation}
where $\|x\|_2 = \left(\sum_i x_i^2\right)^{1/2}$ is the Euclidean norm and we use its square $\|x\|_2^{ 2}$ in the objective for convenience. The coefficients $\epsilon_i \ge 0$ are task-dependent weights. The optimization-based loss ${\rm Loss}_{\mathrm{opt}}$ takes as trainable variables the holographic fields $\Phi$, the potential $V(\Phi)$, and possible interaction terms \cite{Yaraie:2021aa,Hashimoto:2018wm,Hashimoto:2018aa,Song:2020wm,Akutagawa:2020yeo,Park:2022aa,Ahn:2024aa,Ahn:2025aa,Lee:2025aa,Filev:2025aa,Deb:2025aa,Kim:2024aa,Hashimoto:2024aa,Bea:2024aa,Ahn:2024ab}. Also, the optimization-based techniques must be engineered for each instance, with problem specific losses that encode detailed equations of motion (EOM), boundary and horizon conditions (HC), and regularity assumptions for field solutions and potentials. This class of methods is effective when we focus only on a few examples. However, this bespoke design becomes increasingly difficult to apply as the holographic theory ansatz grows more complex because constructing $\rm{Loss}_{\mathrm{opt}}$ demands detailed explicit equations and numerous physical boundary conditions.

In this work, we adopt a data-driven generative AI approach. With a single training procedure, the model can address multiple problems and also unseen cases. Unlike optimization-based methods, this data-driven approach does not require explicit physics information when constructing the AI model and its loss function. As in standard machine learning, the loss function of the data-driven model optimizes the neural network parameters, namely the weights and biases, rather than physical variables.

The Transformer is a suitable model for solving inverse problems. Originally developed for natural language processing, it has become a core architecture across many areas of AI \cite{Vaswani:2017aa}. The model learns correlations between source and target sequences and captures latent patterns. In particular, after a single round of training, it can immediately solve a variety of instances at inference time, which enables it to characterize how solutions change under variations of the data’s control parameters \cite{Kim:2024aa}.

Despite the potential of Transformer to infer dual gravitational theories across a range of gauge theories, their application has been limited by a lack of high-quality training data that enable models to learn the underlying physical structure. Previous work generated and used high-quality datasets, created by randomly selecting coefficients in an analytic form, to train a Transformer, but the trained Transformer achieved lower accuracy on tasks that deviated substantially from the training distribution, presumably because it failed to internalize the underlying physical or mathematical structure \cite{Kim:2024aa}. In this paper, we generate more complex training examples and show that a Transformer trained on such data can capture the relevant physical structure and accurately predict target geometries from the given entanglement entropies.

The Transformer is a sequence-to-sequence model whose core mechanism is attention, which computes correlations across the sequence to predict the next element \cite{Vaswani:2017aa}. The Transformer is generative AI and has become a central mechanism in the AI field. In general, the Transformer consists of the encoder that extracts information from the source sequence into a latent tensor and the decoder that predicts the next target sequence element from the previously generated target sequence and the encoder’s latent tensor. With $x_{1:n}\equiv \{x_1,\ldots,x_n\}$ and $y_{<t}\equiv \{y_1,\ldots,y_{t-1}\}$, we write
\begin{equation}
{\rm Transformer}\left(x_{1:n}, y_{<t}\right) = y_t.
\end{equation}
Given the full source sequence $x_{1:n}$ and the first target element $y_1$, the model predicts $y_2$, and given $x_{1:n}$ and $y_{1:2}$, it predicts $y_3$. Repeating this process yields the entire output sequence $y_{1:m}$.

The Transformer embeds the source and target sequence elements into $d_{model}$-dimensional vectors through an embedding layer and performs computation on these representations. Because the inputs and outputs of encoder and decoder have same shape, the architecture is modular and multiple $N_{enc}$ encoder and $N_{dec}$ decoder blocks can be stacked. Each encoder and decoder block contains multi-head attention and a position-wise feed-forward layer. Multi-head attention uses $h$ heads that compute attention in parallel which improves modeling capacity and efficiency. The feed-forward layer consists of two fully connected layers with a nonlinearity expanding from $d_{model}$ to $d_{ff}$ and projecting back to $d_{model}$ thereby capturing nonlinear structure. During training, dropout with rate $p_{drop}$ helps prevent overfitting by randomly zeroing activations. In this work we use mean squared error as the training loss and Adam as the optimizer to learn the model parameters \cite{Kim:2024aa,Vaswani:2017aa,Kingma:2014aa}. We trained the Transformer in PyTorch and reimplemented it in Mathematica for evaluation. The implementation and test scripts are available at \url{https://github.com/power817/HEE_3D.git}, where the code can be downloaded and exercised.

\section{Review of holographic entanglement entropy}
We introduce a Transformer trained on entanglement entropy from boundary data to predict the dual geometry in AdS$_3$. The target AdS$_3$ metric ansatz is
\begin{equation}
\label{eq:metric}
ds^2=\frac{L^2}{z^2}\left[-f(z)dt^2+\frac{dz^2}{f(z)}+dx^2\right],
\end{equation}
where $z=0$ is the boundary and $L$ is the AdS radius. We focus on black hole geometries, so the blackening function satisfies $f(z_h)=0$. When the boundary is bipartitioned, the holographic entanglement entropy is proportional to the area of the corresponding minimal bulk surface \cite{Ryu:2006ef,Hubeny:2007aa,Casini:2011aa,Headrick:2007aa,Lewkowycz:2013aa,Camps:2013aa,Ryu:2006bv,Ji:2025aa},
\begin{equation}
S(\ell)=\frac{\mathrm{Area}(\gamma_A)}{4G_N},
\end{equation}
where $\gamma_A$ is the minimal surface anchored to the boundary regions.

For a width $\ell$, the time-independent holographic entanglement entropy of the geometry is given by the Ryu--Takayanagi (RT) formula,
\begin{equation}
S(\ell)=\frac{ L}{4 G_N}\int_{-\ell/2}^{\ell/2} dx
\frac{1}{z}\sqrt{1+\frac{(z')^{2}}{f(z)}},\label{eq:Sl}
\end{equation}
where $z'\equiv dz/dx$. Translational symmetry along the $x$-axis implies a conserved quantity $H$,
\begin{equation}
H=-\frac{L}{4G_N z_t},
\end{equation}
where $z_t$ is the turning point at which $dz/dx=0$. Treating Eq.~\eqref{eq:Sl} as a Lagrangian with $x$ playing the role of “time”, the Hamilton--Jacobi equation relates the derivative of the entanglement entropy with respect to $\ell$ to $H$ \cite{Ji:2025aa,Park:2022aa,Ahn:2024aa},
\begin{equation}
\frac{dS}{d\ell}=-H(z_t).\label{eq:HJ}
\end{equation}
As the turning point approaches the black hole horizon, $z_t\to z_h$, the conserved quantity approaches a constant and $S(\ell)$ grows linearly from the large-$\ell$ limit. One then obtains the simpler integral formulae
\begin{align}
\ell(z_t)&=\int_0^{z_t} dz\frac{2}{\sqrt{\left(z_t/z\right)^{2}-1}}\frac{1}{\sqrt{f(z)}},\label{eq:ellzt}\\
S(z_t)&=\frac{L}{2 G_N} \int_{\epsilon_{\rm UV}}^{z_t} dz
\frac{z_t}{z}\frac{1}{\sqrt{z_t^{2}-z^{2}}}\frac{1}{\sqrt{f(z)}},\label{eq:Szt}
\end{align}
where $\epsilon_{\rm UV}$ is a UV cutoff that regulates the divergence.

Once $f(z)$ in our target metric ansatz is given, the entanglement entropy is evaluated directly from the RT formula above. In particular, the Hamilton--Jacobi equation~\eqref{eq:HJ} makes the inverse problem of entanglement entropy more tractable for our Transformer. Although computing the entropy via Eq.~\eqref{eq:Szt} is straightforward, that expression is written in terms of the turning point $z_t$, which is defined in the bulk geometry. Hence, it is not suitable when the source data are $S(\ell)$ as a function of $\ell$. By contrast, the Hamilton--Jacobi equation allows one to determine black hole horizon size $z_h$ from the large-$\ell$ limit and to convert $S(\ell)$ into $\ell(z_t)$ either numerically or analytically. We therefore train a Transformer to recover $f(z)$ from $\ell(z_t)$. When $S(\ell)$ is given, the trained model then reconstructs the dual geometry that reproduces the observed entanglement entropy.

\section{Data generation}

\begin{figure}[t]
  \centering
  \includegraphics[width=0.48\textwidth]{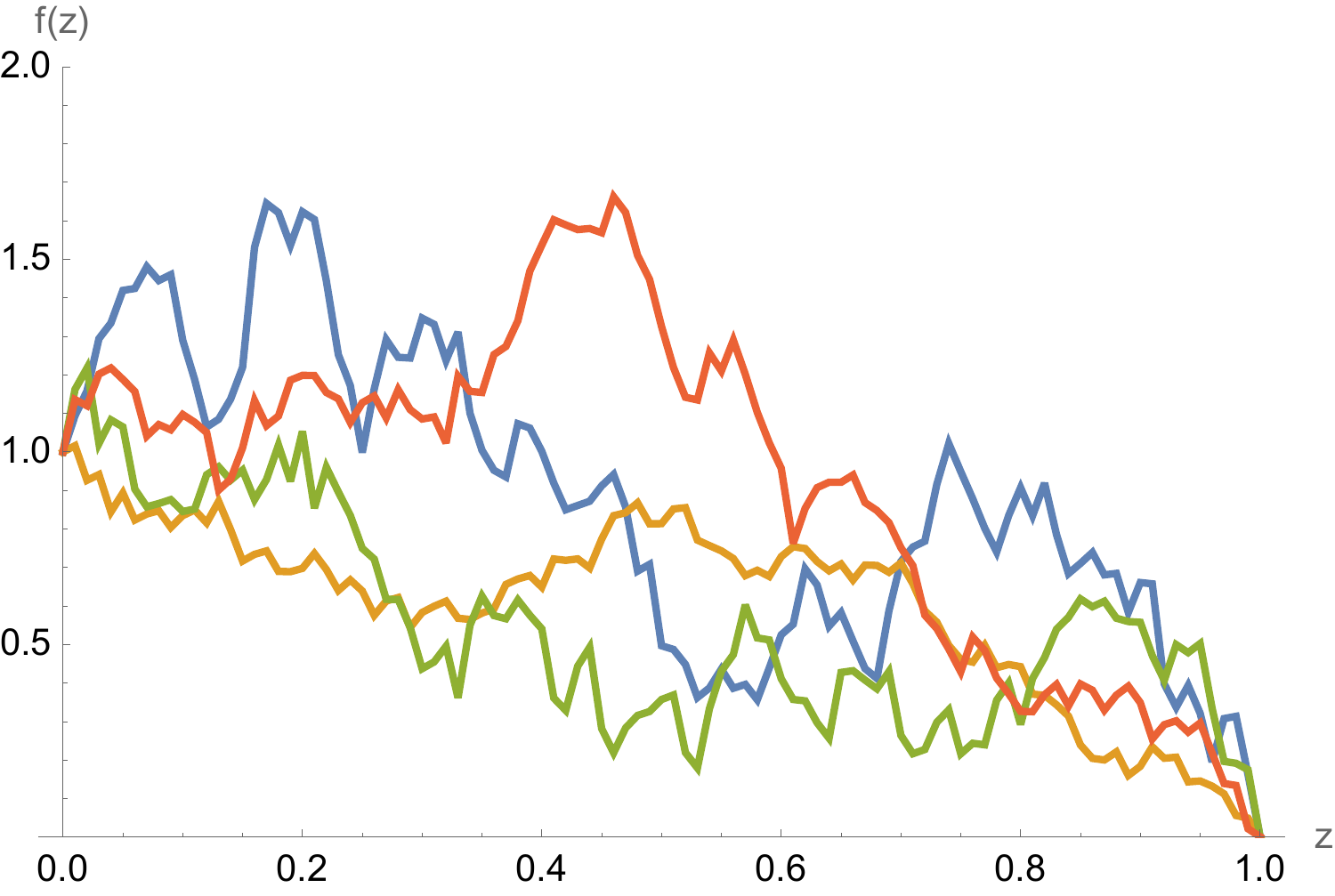}\hfill
  \includegraphics[width=0.48\textwidth]{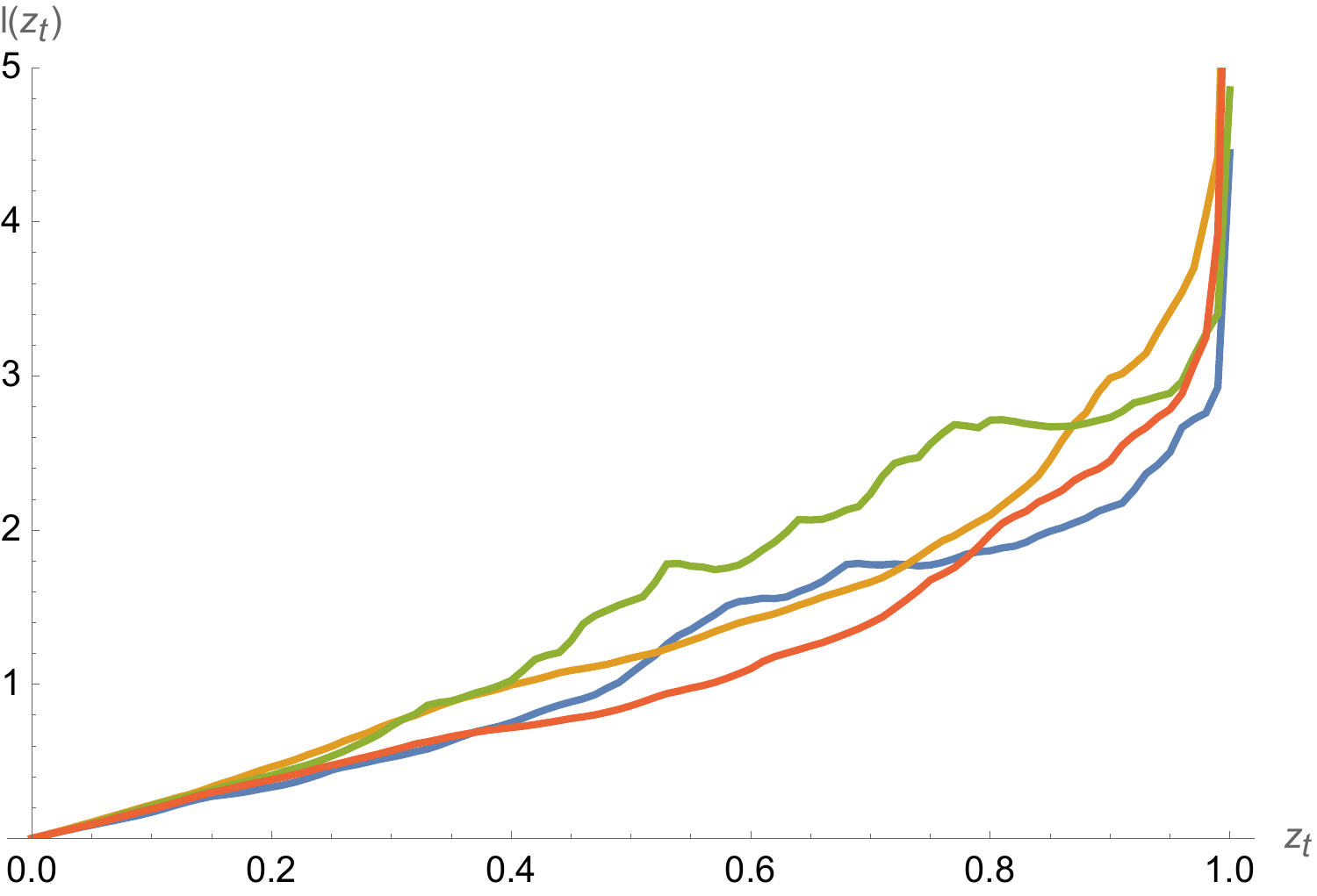}
  \caption{Examples of inputs and targets used in training. Left: blackening functions $f(z)$ with additive white noise $\eta$. Right: the corresponding $\ell(z_t)$ evaluated via Eq.~\eqref{eq:ellzt}. Here $z_h=1$, $z\in[0,1]$ with $\Delta z=0.01$, $\mu=0$, and $\sigma=0.5$. Each figure shows four randomly drawn samples (blue, red, yellow, green), and colors correspond across figures.}
  \label{fig:examples}
\end{figure}

Our aim is to train a Transformer to understand the relevant integral equation well enough to reconstruct the correct bulk dual geometry from previously unseen holographic entanglement data. Achieving this requires a training set that is both large enough and physically meaningful. As discussed above, regardless of the choice of $f(z)$, the time-independent holographic entanglement entropy can be easily computed via the RT formula. Therefore, we can construct the training data by evaluating the RT formula for a range of $f(z)$.

To improve training, we use $\ell(z_t)$ and $f(z)$ rescaled by $z_h$ as source inputs and target outputs. As noted above, the Hamilton--Jacobi formulation allows us to determine a variety of physical observables. In the large-$\ell$ regime, the holographic entanglement entropy grows linearly with $\ell$, whose slope is determined by the horizon. Consequently, although the Transformer is trained only on examples where $\ell(z_t)$ and $f(z)$ are rescaled by $z_h$, the rescale–then–invert procedure at inference time allows the same model to be applied to arbitrary horizon scales $z_h$.

We take the blackening function $f(z)$ to follow a standard BTZ black hole solution added by stochastic white noise \cite{Banados:1992aa},
\begin{equation}
    f(z) = 1 - \left( \frac{z}{z_h}\right)^2 + \eta(z),
\end{equation}
where $\eta(z)$ is a Wiener process in the radial coordinate $z$. In particular, it satisfies
\begin{equation}
    \eta(z+ \Delta z) - \eta(z) \sim {\cal N}\left(\mu \Delta z, \sigma^2 \Delta z\right).
\end{equation}
Here, ${\cal N}$ is a Gaussian distribution with mean $\mu \Delta z$ and variance $\sigma^2 \Delta z$. 

In the forward problem, computing $\ell(z_t)$ via Eq.~\eqref{eq:ellzt} uses only the values of $f(z)$ on the interval $0\le z\le z_t$. Consequently, in the inverse problem, once the geometry is known up to $z\le z_t$, the value of $f$ at $z=z_t+\Delta z$ is constrained by the additional measurement $\ell(z_t+\Delta z)$. Hence the increment $\Delta \ell(z_t)\equiv \ell(z_t+\Delta z)-\ell(z_t)$ is highly sensitive to the local change $\Delta f$ near $z_t$.

By adding $\eta$, we generate $f$ that varies infinitesimally and the corresponding $\ell$ that responds sensitively to it, enabling the Transformer to learn more accurately the hidden inverse RT relation pattern between the two source and target sequences. Once this pattern is learned, the Transformer is better able to make correct inferences on previously unseen data. If the Transformer is trained on datasets that ignore the noise $\eta$, it tends to overconfidently extrapolate subsequent values of $f$ from only the first few generated $f(z)$, which in turn reduces its ability to predict novel dual geometries.

The Transformer is a sequence-to-sequence model, hence both the source and target must be sequences. We therefore sample $z$ at uniform intervals and represent $\ell$ and $f$ as sequences,
\begin{align}
    &{\rm source:} && \{\ell_0,\ldots,\ell_i,\ldots,\ell_N\}, &{\rm target:} && \{f_0,\ldots,f_i,\ldots,f_N\}.
\end{align}
Here $\ell_i=\ell(z_i)$ and $f_i=f(z_i)$, with $\ell_0=0$ and $f_0=1$ by definition and the asymptotically AdS boundary condition. For stability, we admit only samples with $f_i<2$ and $\ell_i<10$ during training. As noted above, to compute $\ell_i$, only the values $f_k$ for $k\le i$ are required. In this setup, the correlation or attention map is approximately the identity matrix,
\begin{equation}
    \langle f_i | \ell_j \rangle \approx  \delta_{ij} .
\end{equation}

In this paper, we take $\mu = 0$ and $\sigma = 0.5$. We sample $z_i$ uniformly from $z=0$ to the horizon (set to $1$) with step $\Delta z=0.01$, yielding vector sequences $\ell_{0:100}$ and $f_{0:100}$ of length 101, where $\ell_i=\ell(z_i)$ and $f_i=f(z_i)$. We generate 100,000 datasets, using $80\%$ for training and $20\%$ for validation. Figure~\ref{fig:examples} illustrates four examples from the generated set. The left figure shows randomly generated blackening functions $f$, and the right figure shows the corresponding $\ell$ computed from each $f$ via Eq.~\eqref{eq:ellzt}.

\section{Model and evaluation}

\begin{figure}[t]
  \centering
  \includegraphics[width=0.95\textwidth]{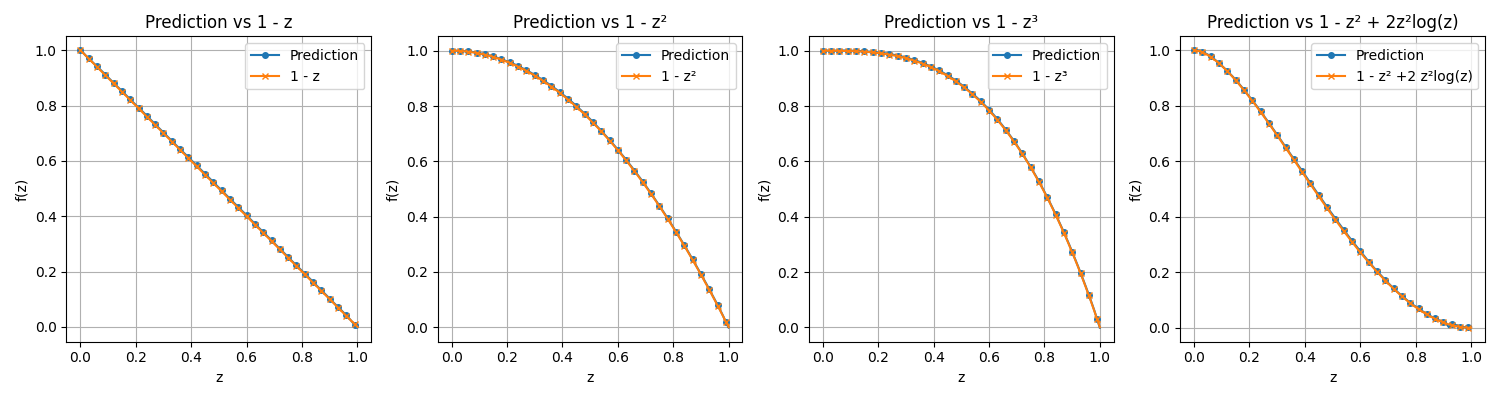}
  \caption{The four plots show outputs from the trained Transformer. The blue curves are the prediction of the AI model, and the orange curves are the true target data.}
  \label{fig:smooth}
\end{figure}

The Transformer is a theory independent sequence model, and once a few hyperparameters are set the model’s size and capacity are largely determined. In this paper, our Transformer has $N_{enc}=N_{dec}=3$, $d_{model}=d_{ff}=512$, and $p_{drop}=0.1$ with a total of about 12M trainable parameters. We built our model using PyTorch’s Transformer module and used the Transformer architecture as originally proposed. After training, the model achieves a training loss of $1.974\times 10^{-5}$ and a validation loss of $1.884\times 10^{-5}$. 

After training, the model reconstructs the blackening function from a variety of entanglement entropy data. The procedure for using the trained Transformer to solve the inverse problem is as follows,
\begin{enumerate}
\item Given analytic or numerical data $S(\ell)$, interpolate the curve and compute the derivative $S'(\ell)$.
\item Use the large-$\ell$ limit to estimate the black hole horizon size. Then, using the Hamilton--Jacobi equation, compute the rescaled $\ell(z_t)$ by the horizon.
\item Convert $\ell(z_t)$ into a sequence compatible with the Transformer's input and feed it to the model. The Transformer generates $f(z)$ autoregressively. From the predicted $f(z)$ compute the entanglement entropy and compare it with the input source to validate the result.
\end{enumerate}

Figure~\ref{fig:smooth} shows the performance on smooth cases. The four plots are the outputs of the trained model. From left to right, the target blackening functions are $1 - z$, $1 - z^2$, $1 - z^3$, and a charged BTZ. The Transformer was trained only on the BTZ black hole solution with metric function $1 - z^2$. Although training used only high-noise data, the trained Transformer accurately predicted smooth dual metric functions that were not seen during training. The ability to predict charged black hole geometries beyond uncharged ones indicates that the models have effectively learned the inverse of the Ryu--Takayanagi formula.

\section{Entanglement entropy variations and dual geometry predictions}
To validate the trained Transformer's performance, we test its ability to predict geometry from entanglement entropy data with unknown dual geometry. The entanglement entropy under consideration has to satisfy specific asymptotic behavior. In the small-$\ell$ limit corresponding to the UV region, it converges to the entanglement entropy of pure AdS \cite{Ryu:2006bv,Ryu:2006ef,Deb:2025aa},
\begin{equation}
    S_{pure}(\ell) = \frac{L}{2G_N}\log\left( \frac{\ell}{\epsilon_{\rm UV}}\right),
\end{equation}
where $L$ denotes the AdS radius, and $G_N$ represents Newton's constant in three dimensions. The factor $L/(2G_N)$ corresponds to the central charge of the dual CFT and is set to one by convention. Because pure AdS has blackening function $f(z) = 1$, the small-$\ell$ asymptotics imply the boundary condition $f(0)=1$.

At finite temperature, the entanglement entropy should have the characteristics of a black hole geometry. In the large $\ell$ limit, it grows linearly with $\ell$, with a slope inversely proportional to the black hole horizon $z_h$, as derived from the Hamilton--Jacobi equation,
\begin{align} 
&S(\ell) \approx \frac{L}{2G_N}\frac{\ell}{2 z_h}.
\end{align}
Using the trained Transformer, we reconstruct the dual geometries from entanglement entropy data in both the finite temperature and zero temperature cases.

The Transformer was trained on inputs with $\max \ell_{0:100} \le 10$ and outputs with $\max f_{0:100} \le 2$. It will still return a prediction even if queries outside these ranges, but its reliability degrades and correct reconstruction is unlikely. Nonetheless, although the trained Transformer was trained on data with $\max f_{0:100}\le 2$ and $\max \ell_{0:100}\le 10$, our tests indicate that it can still make reasonable predictions up to approximately $\max f_{0:100}\approx 2.3$ and $\max \ell_{0:100}\approx 20$. Conceptually, this is analogous to a language model being asked to produce a symbol outside its vocabulary.

\subsection{Case 1: Exponential perturbation}

We consider an entanglement entropy with an exponential perturbation modulated by the parameter $s$,
\begin{equation}
    S(\ell;p) = \frac{L}{2G_N}\log \left[
      \frac{2 z_h}{\epsilon_{\rm UV} (s + 1)}
      \left( e^{\frac{\ell}{2 z_h}} - e^{-s  \frac{\ell}{2 z_h}} \right)
    \right].\label{s_exp}
\end{equation}
This form introduces an asymmetry between the exponential terms, modifying the intermediate-$\ell$ behavior while preserving the asymptotic limits. The parameter $s \ge 0$ does not affect the asymptotics but does influence the shape of the entanglement entropy. In particular, $s=0$ was introduced in \cite{Park:2022aa} as the “unknown” case, representing the simplest form consistent with the asymptotic conditions. 

A homogeneous, gas-like distribution of multiple $p$-branes can be described macroscopically by an effective stress–energy tensor, and in holography the bulk blackening function sourced by the $p$-brane takes the form \cite{park2021holographic,Park:2021aa,Chakrabortty:2011aa,Chakrabortty:2016aa,Park:2020aa,Park:2020ab}
\begin{equation}
f(z) = 1- \left(\frac{z}{z_h}\right)^{2-p},
\end{equation}
where $z_h$ is the black hole horizon and the parameter $p$ interpolates between particle-like ($p=0$) and string like ($p=1$) sources. In particular, for Eq.~\eqref{s_exp} with \(s=0.24651\), the entanglement entropy of the $p$-brane gas geometry with $p=1$ agrees numerically within a mean squared error of $1.71\times 10^{-4}$.  

\begin{figure}[t]
  \centering
  \includegraphics[width=0.48\textwidth]{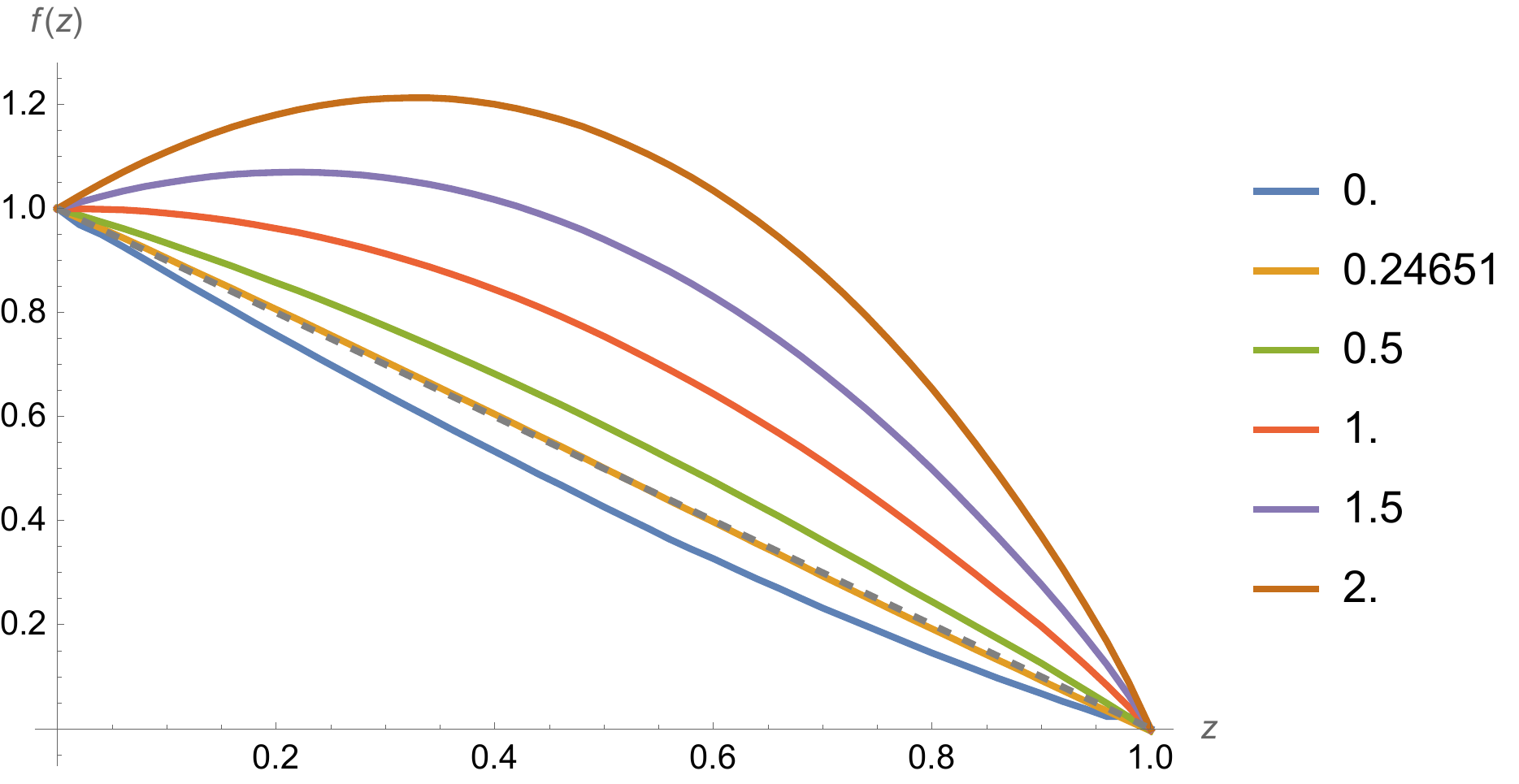}\hfill
  \includegraphics[width=0.48\textwidth]{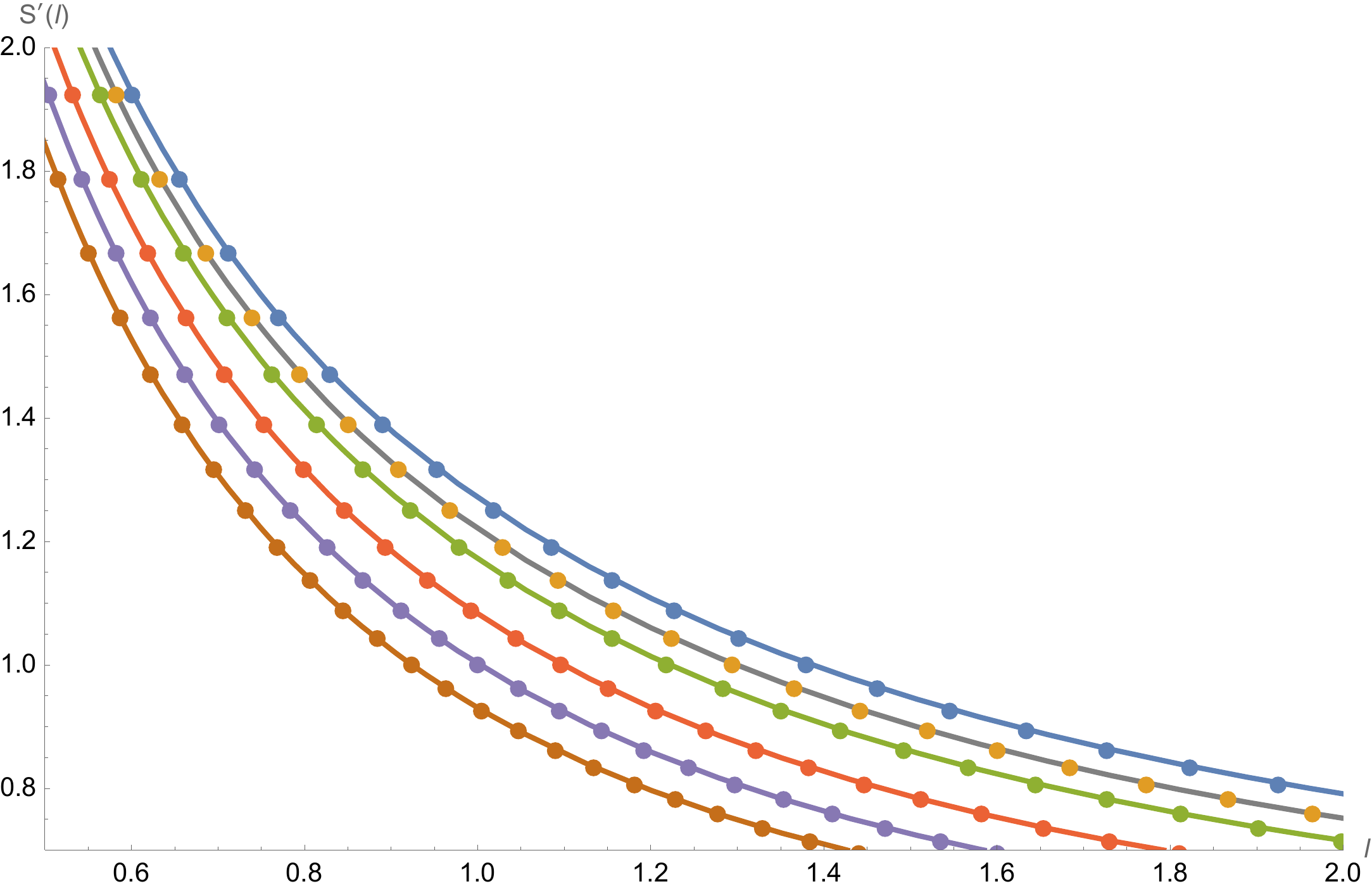}
  \caption{Transformer predictions for the exponential perturbation form of the entanglement entropy. Left: predicted blackening functions $f(z)$ for $s\in\{0,0.24651,0.5,1,1.5,2\}$. Right: solid curves denote the input source $S'(\ell)$, while dotted curves show $S'(\ell)$ recomputed from the predicted $f(z)$.}
  \label{fig:case1}
\end{figure}

In Figure~\ref{fig:case1}, the left figure shows six predictions of $f(z)$ produced by the Transformer, inferred from the entanglement entropy in the right figure, for $s \in [0, 2]$. In the right figure, the dotted curves $S'(\ell)$ are recomputed from the left figure $f(z)$ and coincide with the input entanglement entropy derivatives. In the left figure, the gray dashed line shows $1-z$, the blackening function of the $p=1$ $p$-brane gas geometry. At $s=0.24651$, our prediction closely matches this gray dashed line. In the right figure, the gray solid line likewise denotes the derivative of the entanglement entropy for the $p$-brane gas geometry with $p=1$.

Using 21 blackening functions $f(z;s)$ with $s\in[0,2]$, we employed \texttt{GeneralizedLinearModelFit} (Mathematica) to obtain a bivariate polynomial $f(z;s)$ of total degree at most five. The result is
\begin{align}
    f_{\text{fit}}(z;s) =& 1.00077
    -\left(1.30084+1.31708 s+0.0246912 s^2\right) \frac{z}{z_h}
    \nonumber\\
    & +\left(0.508003-1.65927 s+0.32152 s^2\right) \left(\frac{z}{z_h}\right)^2
    -\left(0.735097+0.486013 s+0.294639 s^2\right) \left(\frac{z}{z_h}\right)^3
    \nonumber\\
    & +\left(0.788506-0.137347 s\right) \left(\frac{z}{z_h}\right)^4
    -0.267425 \left(\frac{z}{z_h}\right)^5 .
\end{align}
To improve stability and interpretability, we set to zero those coefficients with minimal influence according to the coefficient correlation matrix and a sensitivity analysis. Using more than 2,000 evaluation points $\{z,s\}$ sampled from a discrete set $\Omega\subset [0,1]\times[0.5,2]$, the mean squared error is
\begin{equation}
{\rm MSE} = \frac{1}{|\Omega|}\sum_{\{z,s\}\in\Omega} \left(f_{\rm fit}(z;s)-f(z;s)\right)^{2}
= 9.43\times 10^{-7},
\end{equation}
where $f(z;s)$ denotes the blackening function predicted by the Transformer. This MSE is lower than that of the unpruned fitting ansatz. For $s=0.24651$, where Eq.~\eqref{s_exp} most closely matches the $p=1$ $p$-brane gas entanglement entropy, the best-fit blackening function is $f(z;s)\approx 1-z$.

\subsection{Case 2: Hyperbolic-tangent modification}

The entanglement entropy is
\begin{equation}
    S(\ell;s) = \frac{L}{2G_N}\log \left[
      \frac{2 z_h}{\epsilon_{\rm UV}  s} 
      \tanh \left( \frac{s \ell}{2 z_h} \right) 
      e^{\ell/(2 z_h)}
    \right].
\end{equation}
The $\tanh$ factor damps the exponential growth at intermediate scales, with $s$ controlling the transition from UV to IR behavior.

The Transformer produced valid metric predictions for $0.5 \le s \le 3$. This is consistent with the fact that the maximum value of $f_i$ and $\ell_i$ in the training data is 2 and 10. The entanglement entropy for $s>3$ demands $f_i$ values beyond the maximum value, leading the model to incorrect inferences. In the small-$s$ regime, the entropy can be expanded as
\begin{equation}
    S(\ell;s) = \frac{L}{2G_N}\left(\log \left(\frac{\ell  e^{\ell/(2 z_h)}}{\epsilon_{\rm UV}}\right)
    - \frac{s^{2}}{12 z_h^{2}}  \ell^{2} + {\cal O} \left(s^{3}\right)\right).
\end{equation}
In the large-$\ell$ regime there is, in addition to the linear term $\ell/(2 z_h)$, a divergent $\log \ell$ contribution,
so the derivative of the entanglement entropy decays only slowly. Consequently, the required $\ell$ range often exceeds the training bound $\max\ell_{0:100}=10$, pushing the Transformer outside its reliable extrapolation regime. 

\begin{figure}[t]
  \centering
  \includegraphics[width=0.48\textwidth]{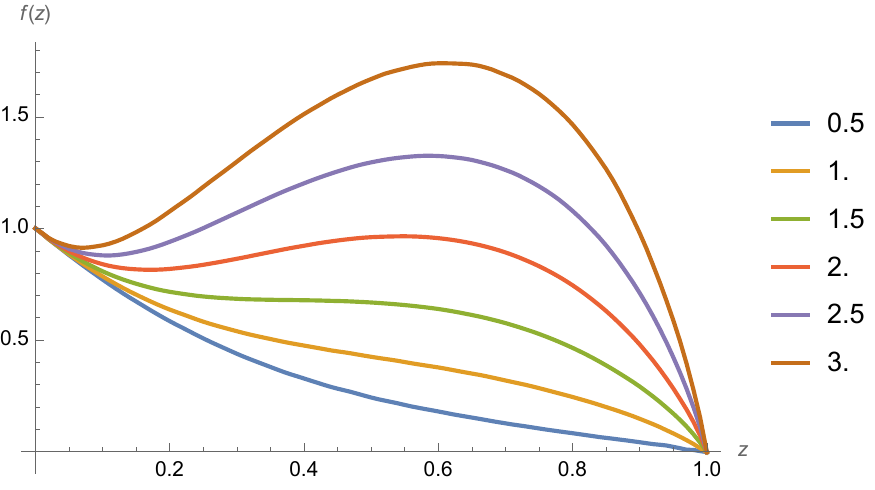}\hfill
  \includegraphics[width=0.48\textwidth]{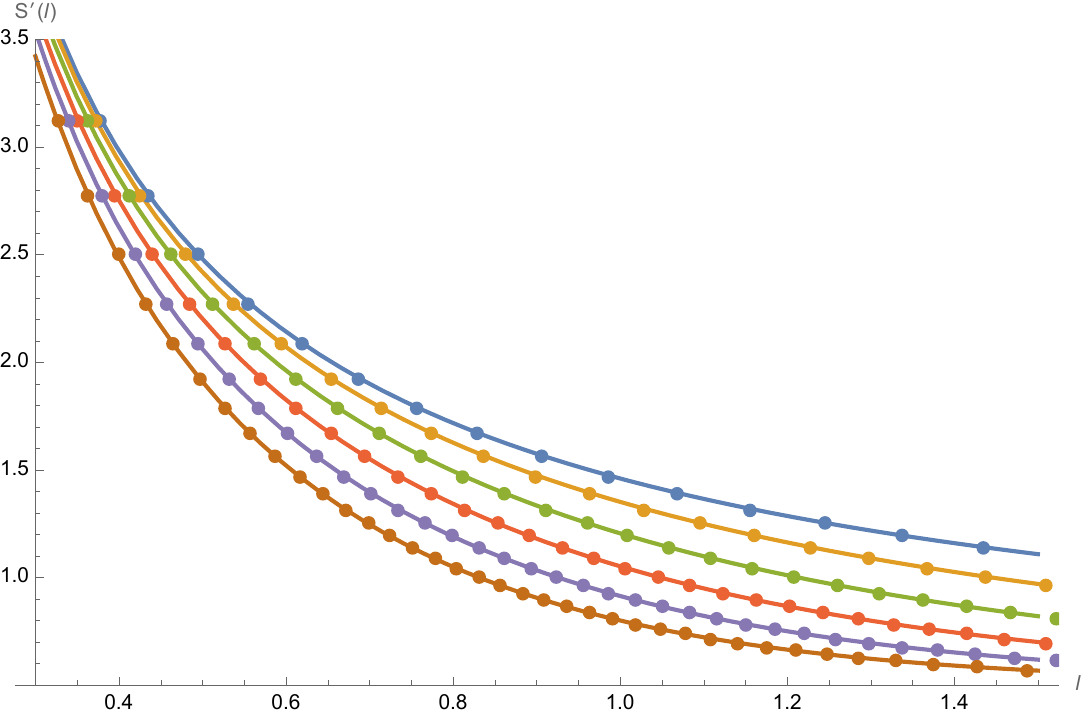}
  \caption{Left: predicted blackening functions $f(z)$ for $s\in\{0.5,1,1.5,2,2.5,3\}$. Right: solid curves denote the input source $S'(\ell)$, while dotted curves show $S'(\ell)$ recomputed from the predicted $f(z)$.}
  \label{fig:case2}
\end{figure}

In Figure~\ref{fig:case2}, the left figure presents predicted blackening functions $f(z)$ for the hyperbolic–tangent modification of the entanglement entropy with $s\in[0.5,3]$ by the trained Transformer. The right figure represents that solid curves show the derivatives $S'(\ell)$ of the entanglement entropies for the same $s$ values, while dotted curves show $S'(\ell)$ recomputed from the predicted $f(z)$. The two sets coincide, confirming consistency. 

Using 26 blackening functions with $s\in[0,3]$ and the same fitting method as before, we fit a bivariate polynomial in $\{z,s\}$ of total degree at most six shown as follows,
\begin{align}
    f_{\rm fit}(z;s) = &1.00002
    -\left(1.96978+0.740627 s-0.202323 s^2\right) \frac{z}{z_h}
    \nonumber\\
    & +\left(4.67945-8.05155 s-0.332988 s^2+0.0885572 s^3\right) \left(\frac{z}{z_h}\right)^2
    \nonumber\\
    & +\left(20.9891-20.3303 s+0.0985536 s^3\right)\left(\frac{z}{z_h}\right)^3
    \nonumber\\
    & -\left(29.1723-20.6611 s+0.569248 s^2\right) \left(\frac{z}{z_h}\right)^4
    + \left(17.3229 - 7.60371 s \right)\left(\frac{z}{z_h}\right)^5
    - 3.49971 \left(\frac{z}{z_h}\right)^6 .
\end{align}
Evaluated on more than 2,000 points drawn from the target domain, the mean squared error is ${\rm MSE}=6.78 \times 10^{-6}$.

\subsection{Case 3: Power interpolation}

This case uses a power weighted interpolation,
\begin{equation}
    S(\ell) = \frac{L}{2G_N}\log\left[ \frac{2 z_h}{\epsilon_{\rm UV}} \frac{\frac{\ell}{2 z_h} + e^{\frac{\ell}{2 z_h}} \left(\frac{\ell}{2 z_h}\right)^s}{1 + \left(\frac{\ell}{2 z_h}\right)^s} \right].
\end{equation}
Although this entanglement entropy form also satisfies both asymptotic regimes for $s > 1$, it violates the UV behavior when $s\leq 1$. The slope of the entanglement entropy curve at finite $\ell$ depends on the parameter $s$. Notably, beyond a certain value of $s$, the instantaneous slope $S'(\ell)$ drops below 1. In our context, a unit slope signals near–black hole behavior and indicates the emergence of a new black hole horizon.

\begin{figure}[t]
  \centering
  \includegraphics[width=0.48\textwidth]{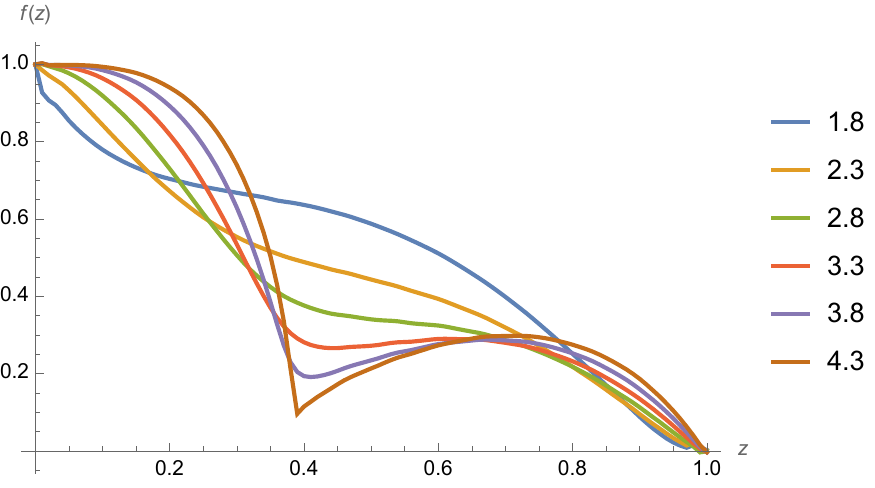}\hfill
  \includegraphics[width=0.48\textwidth]{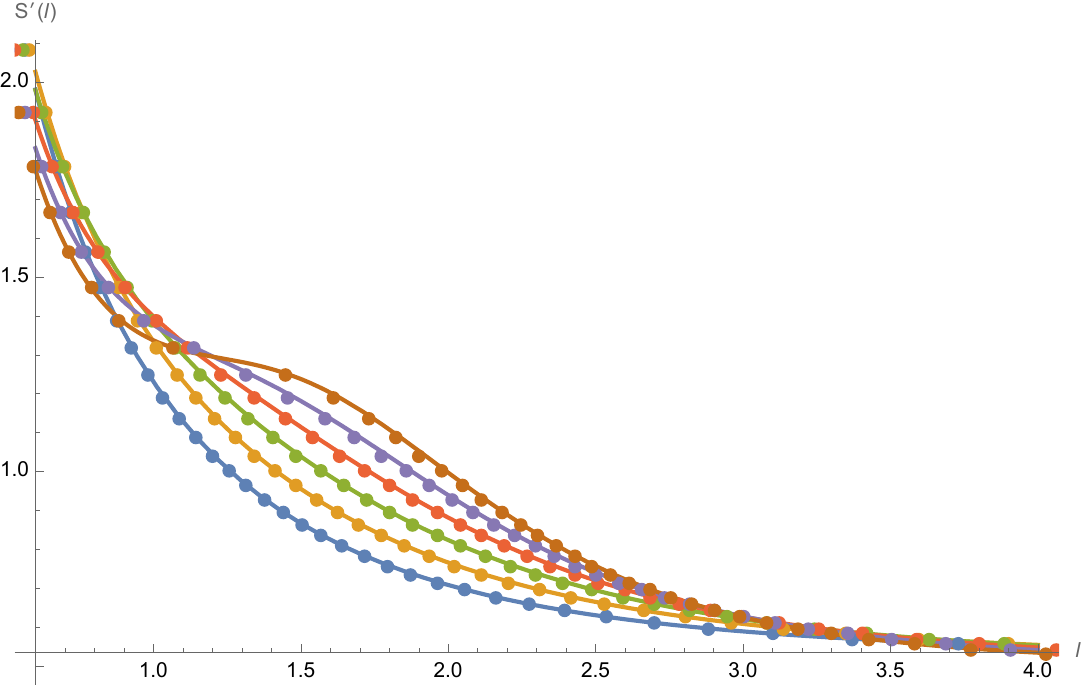}
  \caption{Left: predicted blackening functions $f(z)$ for $s\in\{1.8, 2.3, 2.8, 3.3, 3.8, 4.3\}$. Right: solid curves denote the input source $S'(\ell)$, while dotted curves show $S'(\ell)$ recomputed from the predicted $f(z)$.}
  \label{fig:case3}
\end{figure}

Within the range where the trained Transformer makes reliable inferences, we therefore select  $s \in [1.8,  4.3]$, as shown in Figure~\ref{fig:case3}. Based on the Transformer’s predictions, for $s>4.3$ the instantaneous slope $S'(\ell)$ in the finite $\ell$ satisfies $S'(\ell)<1$, consistent with the appearance of an additional black hole horizon.

The function $ f(z, s) $ is smooth for small $ s $, but develops sharp spikes or singular behavior as $ s $ exceeds a critical value. We were unable to find a suitable fitting function using the same approach as before. Despite applying the previous method, the model fails to capture the sharp, parameter-dependent behavior of $f(z; s)$.

\subsection{Case 4: Periodic boundary entanglement entropy}

Although the Transformer was only trained on datasets satisfying the black hole condition $f(z_h)=0$, from a mathematical standpoint $\ell(z_t)$ can be computed for any regular function $f$. If the model has understood the underlying integral equation, it should generalize beyond the training distribution. Despite being trained only on black hole geometries, it would be able to produce sensible predictions for non black hole cases as well. We now test this using an entanglement entropy that is periodic in the $x$ direction and is specified as follows \cite{Ryu:2006bv,Ryu:2006ef,Deb:2025aa},
\begin{equation}
    S(\ell) = \frac{L}{2G_N}\log\left[ \frac{2 s}{\epsilon_{\rm UV}} \sin\left( \frac{\ell}{2 s} \right) \right].
\end{equation}

\begin{figure}[t]
  \centering
  \includegraphics[width=0.48\textwidth]{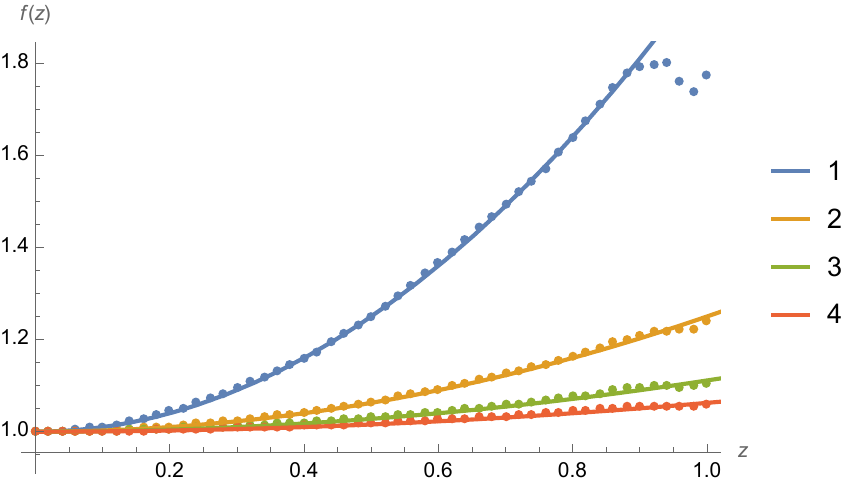}
  \caption{The dotted curves denote predicted blackening functions $f(z)$ for $s\in\{1,2,3,4\}$, and solid curves show the true blackening function $1+(z/s)^2$.}
  \label{fig:case4}
\end{figure}

Figure~\ref{fig:case4} shows a comparison between the dual geometry predicted by the Transformer and the true blackening function $f(z)$, both constructed from periodic boundary entanglement entropy. Near $z=1$ the Transformer's predictions become unstable. Nonetheless, its predictions generally align well with the true data. Although the Transformer was only trained on cases with $f_{100}=0$, it clearly learned the integral equation well enough to generalize beyond the training distribution.

\section{Discussion}

This study shows that a Transformer trained on a holographic dataset with white noise learns the inverse mapping of the Ryu--Takayanagi formula in three dimensions with high accuracy and can predict unseen geometries. Given $\ell(z_t)$ as input, the trained Transformer outputs the corresponding blackening function $f(z)$ not only for noisy data but also for smooth black holes and even for horizonless geometries. Although the training set contained only BTZ solutions satisfying $f(z_h)=0$ with additive white noise, the trained Transformer performs well on cases not seen during training. This indicates that it has learned the functional inverse relation of the RT integral.

From multiple case studies, we identify several properties of the trained Transformer. Reliability holds within the training range $\ell_i \le 20$ and $f_i \le 2.3$. Outside this range the prediction quality degrades sharply, as seen for $s>3$ in the hyperbolic tangent modification cases and for $s>4.3$ in the power interpolation cases. This behavior is expected for sequence models and, in physical terms, corresponds to probing temperature regimes not covered by the training data.

Within the valid regime the model produces physically consistent results. For the exponential and hyperbolic tangent cases, polynomial fits to the predicted $f(z;s)$ achieve mean squared errors in the range $10^{-6}$ to $10^{-7}$. The fitted coefficients vary smoothly with the parameter $s$, which shows that the model captures systematic geometric deformations. In the power interpolation cases the polynomial fit fails at large $s$, coinciding with the emergence of sharp structure in $f(z)$ and with points where $S'(\ell)$ drops below the black hole condition.

Even though the training data included only black hole geometries with $f(z_h)=0$, the trained Transformer predicts a horizonless dual geometry from the test in the periodic boundary entropy case. An instability appears near $z=1$, yet the overall profile is accurate. This supports the view that the Transformer has understood the core integral relation beyond the assumptions used during training.

A data-driven approach, after a single training, can handle diverse unknown examples and does not require constructing loss functions or explicit boundary conditions. The attention mechanism captures the nonlocal correspondence between boundary size $\ell(z_t)$ and dual geometry $f(z)$. Adding Wiener noise was crucial for capturing the structure of the integral relation. Without noise the Transformer fails to generalize. Noise teaches local differential sensitivity, which is essential for accurate inverse mapping.

The present framework has limitations. It targets AdS$_3$ and a single metric function $f(z)$. Extending to higher dimensions, spherical or irregular regions, or time dependent entropies will require larger and more diverse datasets and possibly modified architectures. While the geometry is reconstructed accurately, extracting physical quantities such as temperature or charge still needs post processing such as fitting.

Although the present work focuses exclusively on AdS$_3$ backgrounds, it would be interesting in future studies to investigate whether a Transformer trained on higher-dimensional entanglement data can learn the corresponding inverse RT formula in more general settings. In particular, top-down constructions such as those studied by \cite{Jang:2017aa} as well as the smooth, horizonless LLM geometries of \cite{Lin:2004aa} offer explicit warp factors in four and higher dimensions that may provide suitable data for training a higher-dimensional model. These examples suggest that a Transformer for higher-dimensional geometries might also learn the inverse RT relation in those settings, though confirming this would require separate investigation.

\section*{Acknowledgments}
This work was partly supported by the KIAS Individual Grant (AP103401) at the Korea Institute for Advanced Study.
We thank the Center for Advanced Computation in KIAS for providing computing resources.

\bibliography{ref}

@article{Lin:2004aa,
	author = {Hai Lin and Oleg Lunin and Juan Maldacena},
	doi = {10.1088/1126-6708/2004/10/025},
	eprint = {hep-th/0409174},
	journal = {JHEP },
	pages = {025},
	title = {Bubbling AdS space and 1/2 BPS geometries},
	url = {https://arxiv.org/pdf/hep-th/0409174.pdf},
	volume = {0410},
	year = {2004}}

@article{Jang:2017aa,
	author = {Dongmin Jang and Yoonbai Kim and O-Kab Kwon and D. D. Tolla},
	eprint = {1712.09101},
	month = {12},
	title = {Gravity from Entanglement and RG Flow in a Top-down Approach},
	url = {https://arxiv.org/pdf/1712.09101.pdf},
	year = {2017}}

@article{Lu:2025ab,
	author = {Zhenkang Lu and Cheng Ran and Shao-feng Wu},
	eprint = {2507.13306},
	month = {07},
	title = {The Algebraic Structure Underlying Pole-Skipping Points},
	url = {https://arxiv.org/pdf/2507.13306.pdf},
	year = {2025}}

@article{Lu:2025aa,
	author = {Zhenkang Lu and Cheng Ran and Shao-feng Wu},
	eprint = {2506.12890},
	month = {06},
	title = {Bulk Spacetime Encoding via Boundary Ambiguities},
	url = {https://arxiv.org/pdf/2506.12890.pdf},
	year = {2025}}

@article{Park:2020ab,
	author = {Chanyong Park},
	doi = {10.1140/epjc/s10052-021-09308-0},
	eprint = {2011.13555},
	month = {11},
	title = {Time-dependent quantum correlations in two-dimensional expanding spacetime},
	url = {https://arxiv.org/pdf/2011.13555.pdf},
	year = {2020}}

@article{Park:2020aa,
	author = {Chanyong Park},
	doi = {10.1103/PhysRevD.101.126006},
	eprint = {2004.08020},
	journal = {Phys. Rev. D},
	pages = {126006},
	title = {Time Evolution of Entanglement Entropy in Holographic FLRW Cosmologies},
	url = {https://arxiv.org/pdf/2004.08020.pdf},
	volume = {101},
	year = {2020}}

@article{Chakrabortty:2016aa,
	author = {Shankhadeep Chakrabortty and Tanay K. Dey},
	eprint = {1602.04761},
	month = {02},
	title = {Back reaction effects on the dynamics of heavy probes in heavy quark cloud},
	url = {https://arxiv.org/pdf/1602.04761.pdf},
	year = {2016}}

@article{Chakrabortty:2011aa,
	author = {Shankhadeep Chakrabortty},
	doi = {10.1016/j.physletb.2011.09.112},
	eprint = {1108.0165},
	month = {08},
	title = {Dissipative force on an external quark in heavy quark cloud},
	url = {https://arxiv.org/pdf/1108.0165.pdf},
	year = {2011}}

@article{Park:2021aa,
	author = {Chanyong Park and Jung Hun Lee},
	doi = {10.1007/s40042-022-00669-7},
	eprint = {2102.06097},
	month = {02},
	title = {Quantum correlation in quark-gluon medium},
	url = {https://arxiv.org/pdf/2102.06097.pdf},
	year = {2021}}

@article{Yaraie:2021aa,
	author = {Emad Yaraie and Hossein Ghaffarnejad and Mohammad Farsam},
	doi = {https://doi.org/10.22128/ijaa.2023.694.1150},
	eprint = {2108.07161},
	journal = {Iranian Journal of Astronomy and Astrophysics 2023},
	month = {08},
	title = {Physics-informed deep learning for three dimensional black holes},
	url = {https://arxiv.org/pdf/2108.07161.pdf},
	year = {2021}}

@article{Champion:2019aa,
	author = {Kathleen Champion and Peng Zheng and Aleksandr Y. Aravkin and Steven L. Brunton and J. Nathan Kutz},
	eprint = {1906.10612},
	month = {06},
	title = {A unified sparse optimization framework to learn parsimonious physics-informed models from data},
	url = {https://arxiv.org/pdf/1906.10612.pdf},
	year = {2019}}

@article{Raissi:2017ab,
	author = {Maziar Raissi and Paris Perdikaris and George Em Karniadakis},
	eprint = {1711.10566},
	month = {11},
	title = {Physics Informed Deep Learning (Part II): Data-driven Discovery of Nonlinear Partial Differential Equations},
	url = {https://arxiv.org/pdf/1711.10566.pdf},
	year = {2017}}

@article{Ahn:2024ab,
	author = {Byoungjoon Ahn and Hyun-Sik Jeong and Keun-Young Kim and Kwan Yun},
	doi = {10.1007/JHEP03(2024)141},
	eprint = {2401.00939},
	journal = {J. High Energ. Phys.},
	pages = {141},
	title = {Deep learning bulk spacetime from boundary optical conductivity},
	url = {https://arxiv.org/pdf/2401.00939.pdf},
	volume = {2024},
	year = {2024}}

@article{Bea:2024aa,
	author = {Yago Bea and Raul Jimenez and David Mateos and Shuheng Liu and Pavlos Protopapas and Pedro Taranc{\'o}n-{\'A}lvarez and Pablo Tejerina-P{\'e}rez},
	doi = {10.1007/JHEP07(2024)087},
	eprint = {2403.14763},
	month = {03},
	title = {Gravitational Duals from Equations of State},
	url = {https://arxiv.org/pdf/2403.14763.pdf},
	year = {2024}}

@article{Hashimoto:2024aa,
	author = {Koji Hashimoto and Koshiro Matsuo and Masaki Murata and Gakuto Ogiwara and Daichi Takeda},
	eprint = {2411.16052},
	month = {11},
	title = {Machine-learning emergent spacetime from linear response in future tabletop quantum gravity experiments},
	url = {https://arxiv.org/pdf/2411.16052.pdf},
	year = {2024}}

@article{Bao:2019aa,
	author = {Ning Bao and ChunJun Cao and Sebastian Fischetti and Cynthia Keeler},
	doi = {10.1088/1361-6382/ab377f},
	eprint = {1904.04834},
	month = {04},
	title = {Towards Bulk Metric Reconstruction from Extremal Area Variations},
	url = {https://arxiv.org/pdf/1904.04834.pdf},
	year = {2019}}

@article{Bilson:2011aa,
	author = {Samuel Bilson},
	doi = {10.1007/JHEP02(2011)050},
	eprint = {1012.1812},
	journal = {JHEP},
	pages = {050},
	title = {Extracting Spacetimes using the AdS/CFT Conjecture: Part II},
	url = {https://arxiv.org/pdf/1012.1812.pdf},
	volume = {1102},
	year = {2011}}

@article{Bilson:2008aa,
	author = {Samuel Bilson},
	doi = {10.1088/1126-6708/2008/08/073},
	eprint = {0807.3695},
	journal = {JHEP},
	pages = {073},
	title = {Extracting spacetimes using the AdS/CFT conjecture},
	url = {https://arxiv.org/pdf/0807.3695.pdf},
	volume = {0808},
	year = {2008}}

@article{Hamilton:2006aa,
	author = {Alex Hamilton and Daniel Kabat and Gilad Lifschytz and David A. Lowe},
	doi = {10.1103/PhysRevD.74.066009},
	eprint = {hep-th/0606141},
	journal = {Phys.Rev.D},
	pages = {066009},
	title = {Holographic representation of local bulk operators},
	url = {https://arxiv.org/pdf/hep-th/0606141.pdf},
	volume = {74},
	year = {2006}}

@article{Ramallo:2013aa,
	author = {Alfonso V. Ramallo},
	eprint = {1310.4319},
	month = {10},
	title = {Introduction to the AdS/CFT correspondence},
	url = {https://arxiv.org/pdf/1310.4319.pdf},
	year = {2013}}

@article{Filev:2025aa,
	author = {Veselin G. Filev},
	eprint = {2506.20115},
	month = {06},
	title = {Holographic flavour and neural networks},
	url = {https://arxiv.org/pdf/2506.20115.pdf},
	year = {2025}}

@article{Jokela:2025aa,
	author = {Niko Jokela and Tony Liimatainen and Miika Sarkkinen and Leo Tzou},
	eprint = {2504.07016},
	month = {04},
	title = {Bulk metric reconstruction from entanglement data via minimal surface area variations},
	url = {https://arxiv.org/pdf/2504.07016.pdf},
	year = {2025}}

@article{Lee:2025aa,
	author = {Donghee Lee and Hye-Sung Lee and Jaeok Yi},
	eprint = {2503.08827},
	month = {03},
	title = {Synaptic Field Theory for Neural Networks},
	url = {https://arxiv.org/pdf/2503.08827.pdf},
	year = {2025}}

@article{Ahn:2025aa,
	author = {Byoungjoon Ahn and Hyun-Sik Jeong and Chang-Woo Ji and Keun-Young Kim and Kwan Yun},
	eprint = {2502.10245},
	month = {02},
	title = {Deep learning-based holography for T-linear resistivity},
	url = {https://arxiv.org/pdf/2502.10245.pdf},
	year = {2025}}

@article{Hashimoto:2018aa,
	author = {Koji Hashimoto and Sotaro Sugishita and Akinori Tanaka and Akio Tomiya},
	doi = {10.1103/PhysRevD.98.106014},
	eprint = {1809.10536},
	journal = {Phys. Rev. D},
	pages = {106014},
	title = {Deep Learning and Holographic QCD},
	url = {https://arxiv.org/pdf/1809.10536.pdf},
	volume = {98},
	year = {2018}}

@article{Deb:2025aa,
	author = {Anirudh Deb and Yaman Sanghavi},
	eprint = {2509.25311},
	journal = {YITP-SB-2025-18},
	month = {09},
	title = {Aspects of holographic entanglement using physics-informed-neural-networks},
	url = {https://arxiv.org/pdf/2509.25311.pdf},
	year = {2025}}

@article{Park:2022aa,
	author = {Chanyong Park and Chi-Ok Hwang and Kyungchan Cho and Se-Jin Kim},
	doi = {https://doi.org/10.1103/PhysRevD.106.106017},
	eprint = {2205.04445},
	month = {05},
	title = {Dual Geometry of Entanglement Entropy via Deep Learning},
	url = {https://arxiv.org/pdf/2205.04445.pdf},
	year = {2022}}

@article{Kim:2024aa,
	author = {Sejin Kim and Kyung Kiu Kim and Yunseok Seo},
	eprint = {2410.06523},
	month = {10},
	title = {Phase Diagram from Nonlinear Interaction between Superconducting Order and Density: Toward Data-Based Holographic Superconductor},
	url = {https://arxiv.org/pdf/2410.06523.pdf},
	year = {2024}}

@article{Park:2023aa,
	author = {Chanyong Park and Sejin Kim and Jung Hun Lee},
	eprint = {2311.01724},
	month = {11},
	title = {Holography Transformer},
	url = {https://arxiv.org/pdf/2311.01724.pdf},
	year = {2023}}

@article{Ahn:2024aa,
	author = {Byoungjoon Ahn and Hyun-Sik Jeong and Keun-Young Kim and Kwan Yun},
	doi = {https://doi.org/10.1007/JHEP01%282025%29025},
	eprint = {2406.07395},
	journal = {IFT-UAM/CSIC-24-88},
	month = {06},
	title = {Holographic reconstruction of black hole spacetime: machine learning and entanglement entropy},
	url = {https://arxiv.org/pdf/2406.07395.pdf},
	year = {2024}}

@article{Ji:2025aa,
	author = {Xuanting Ji and Xin-Xiang Ju and Ya-Wen Sun and Yuan-Tai Wang and He-Lin Zhou},
	doi = {https://doi.org/10.1007/JHEP09%282025%29081},
	eprint = {2505.08534},
	journal = {JHEP, 2025, 09: 081},
	month = {05},
	title = {Holographic geometry/real-space entanglement correspondence and metric reconstruction},
	url = {https://arxiv.org/pdf/2505.08534.pdf},
	year = {2025}}

@article{Dong:2016aa,
	author = {Xi Dong and Daniel Harlow and Aron C. Wall},
	doi = {https://doi.org/10.1103/PhysRevLett.117.021601},
	eprint = {1601.05416},
	journal = {NSF-KITP-16-005},
	month = {01},
	title = {Reconstruction of Bulk Operators within the Entanglement Wedge in Gauge-Gravity Duality},
	url = {https://arxiv.org/pdf/1601.05416.pdf},
	year = {2016}}

@article{Czech:2015ab,
	author = {Bartlomiej Czech and Lampros Lamprou and Samuel McCandlish and James Sully},
	doi = {10.1007/JHEP07(2016)100},
	eprint = {1512.01548},
	month = {12},
	title = {Tensor Networks from Kinematic Space},
	url = {https://arxiv.org/pdf/1512.01548.pdf},
	year = {2015}}

@article{Camps:2013aa,
	author = {Joan Camps},
	doi = {https://doi.org/10.1007/JHEP03%282014%29070},
	eprint = {1310.6659},
	month = {10},
	title = {Generalized entropy and higher derivative Gravity},
	url = {https://arxiv.org/pdf/1310.6659.pdf},
	year = {2013}}

@article{Lewkowycz:2013aa,
	author = {Aitor Lewkowycz and Juan Maldacena},
	doi = {https://doi.org/10.1007/JHEP08%282013%29090},
	eprint = {1304.4926},
	month = {04},
	title = {Generalized gravitational entropy},
	url = {https://arxiv.org/pdf/1304.4926.pdf},
	year = {2013}}

@article{Headrick:2007aa,
	author = {Matthew Headrick and Tadashi Takayanagi},
	doi = {https://doi.org/10.1103/PhysRevD.76.106013},
	eprint = {0704.3719},
	journal = {SU-ITP-07/08, KUNS-2069},
	month = {04},
	title = {A holographic proof of the strong subadditivity of entanglement entropy},
	url = {https://arxiv.org/pdf/0704.3719.pdf},
	year = {2007}}

@article{Casini:2011aa,
	author = {Horacio Casini and Marina Huerta and Robert C. Myers},
	doi = {https://doi.org/10.1007/JHEP05%282011%29036},
	eprint = {1102.0440},
	journal = {JHEP},
	pages = {036},
	title = {Towards a derivation of holographic entanglement entropy},
	url = {https://arxiv.org/pdf/1102.0440.pdf},
	volume = {1105},
	year = {2011}}

@article{Hubeny:2007aa,
	author = {Veronika E. Hubeny and Mukund Rangamani and Tadashi Takayanagi},
	doi = {https://doi.org/10.1088/1126-6708/2007/07/062},
	eprint = {0705.0016},
	journal = {JHEP},
	pages = {062},
	title = {A Covariant Holographic Entanglement Entropy Proposal},
	url = {https://arxiv.org/pdf/0705.0016.pdf},
	volume = {0707},
	year = {2007}}

@article{Banados:1992aa,
	author = {M{\'a}ximo Ba{\~n}ados and Claudio Teitelboim and Jorge Zanelli},
	doi = {10.1103/PhysRevLett.69.1849},
	eprint = {hep-th/9204099},
	journal = {Phys.Rev.Lett.},
	pages = {1849-1851},
	title = {The Black Hole in Three Dimensional Space Time},
	url = {https://arxiv.org/pdf/hep-th/9204099.pdf},
	volume = {69},
	year = {1992}}

@article{Vaswani:2017aa,
	author = {Ashish Vaswani and Noam Shazeer and Niki Parmar and Jakob Uszkoreit and Llion Jones and Aidan N. Gomez and Lukasz Kaiser and Illia Polosukhin},
	eprint = {1706.03762},
	month = {06},
	title = {Attention Is All You Need},
	url = {https://arxiv.org/pdf/1706.03762.pdf},
	year = {2017}}

@article{Kingma:2014aa,
	author = {Diederik P. Kingma and Jimmy Ba},
	eprint = {1412.6980},
	month = {12},
	title = {Adam: A Method for Stochastic Optimization},
	url = {https://arxiv.org/pdf/1412.6980.pdf},
	year = {2014}}

@article{Raissi:2017aa,
	author = {Maziar Raissi and Paris Perdikaris and George Em Karniadakis},
	eprint = {1711.10561},
	month = {11},
	title = {Physics Informed Deep Learning (Part I): Data-driven Solutions of Nonlinear Partial Differential Equations},
	url = {https://arxiv.org/pdf/1711.10561.pdf},
	year = {2017}}

@article{Witten:1998aa,
	author = {Edward Witten},
	eprint = {hep-th/9802150},
	journal = {Adv.Theor.Math.Phys.},
	pages = {253-291},
	title = {Anti De Sitter Space And Holography},
	url = {https://arxiv.org/pdf/hep-th/9802150.pdf},
	volume = {2},
	year = {1998}}

@article{Akutagawa:2020yeo,
	archiveprefix = {arXiv},
	author = {Akutagawa, Tetsuya and Hashimoto, Koji and Sumimoto, Takayuki},
	doi = {10.1103/PhysRevD.102.026020},
	eprint = {2005.02636},
	journal = {Phys. Rev. D},
	number = {2},
	pages = {026020},
	primaryclass = {hep-th},
	reportnumber = {OU-HET 1058},
	title = {{Deep Learning and AdS/QCD}},
	volume = {102},
	year = {2020}}

@article{park2021holographic,
	archiveprefix = {arXiv},
	author = {Chanyong Park},
	eprint = {2106.05500},
	primaryclass = {hep-th},
	title = {{Holographic time-dependent entanglement entropy in $p$-brane gas geometries}},
	url = {https://arxiv.org/pdf/2106.05500.pdf},
	year = {2021}}

@article{Hashimoto:2018wm,
	author = {Koji Hashimoto and Sotaro Sugishita and Akinori Tanaka and Akio Tomiya},
	doi = {10.1103/PhysRevD.98.046019},
	eprint = {1802.08313},
	journal = {Phys. Rev. D},
	pages = {046019},
	title = {{Deep Learning and AdS/CFT}},
	url = {https://arxiv.org/pdf/1802.08313.pdf},
	volume = {98},
	year = {2018}}

@article{Song:2020wm,
	author = {Mugeon Song and Maverick S. H. Oh and Yongjun Ahn and Keun-Young Kim},
	doi = {10.1088/1674-1137/abfc36},
	eprint = {2011.13726},
	month = {11},
	title = {{AdS/Deep-Learning made easy: simple examples}},
	url = {https://arxiv.org/pdf/2011.13726.pdf},
	year = {2020}}

@article{Ryu:2006ef,
	archiveprefix = {arXiv},
	author = {Ryu, Shinsei and Takayanagi, Tadashi},
	doi = {10.1088/1126-6708/2006/08/045},
	eprint = {hep-th/0605073},
	journal = {JHEP},
	pages = {045},
	primaryclass = {hep-th},
	reportnumber = {NSF-KITP-06-31, KUNS-2021},
	slaccitation = {%%CITATION = HEP-TH/0605073;%%},
	title = {{Aspects of Holographic Entanglement Entropy}},
	volume = {08},
	year = {2006}}

@article{Ryu:2006bv,
	archiveprefix = {arXiv},
	author = {Ryu, Shinsei and Takayanagi, Tadashi},
	doi = {10.1103/PhysRevLett.96.181602},
	eprint = {hep-th/0603001},
	journal = {Phys. Rev. Lett.},
	pages = {181602},
	primaryclass = {hep-th},
	reportnumber = {NSF-KITP-06-11},
	slaccitation = {%%CITATION = HEP-TH/0603001;%%},
	title = {{Holographic derivation of entanglement entropy from AdS/CFT}},
	volume = {96},
	year = {2006}}

@article{Maldacena:1997re,
	archiveprefix = {arXiv},
	author = {Maldacena, Juan Martin},
	doi = {10.1023/A:1026654312961, 10.4310/ATMP.1998.v2.n2.a1},
	eprint = {hep-th/9711200},
	journal = {Int. J. Theor. Phys.},
	note = {[Adv. Theor. Math. Phys.2,231(1998)]},
	pages = {1113-1133},
	primaryclass = {hep-th},
	reportnumber = {HUTP-97-A097, HUTP-98-A097},
	slaccitation = {%%CITATION = HEP-TH/9711200;%%},
	title = {{The Large N limit of superconformal field theories and supergravity}},
	volume = {38},
	year = {1999}}
\bibliographystyle{unsrtnat}

\end{document}